\documentclass[%
 reprint,
superscriptaddress,
preprintnumbers,
 amsmath,amssymb,
 aps,
]{revtex4-2}

\usepackage{graphicx}% Include figure files
\usepackage{dcolumn}% Align table columns on decimal point
\usepackage{bm}% bold math
\usepackage{hyperref}% add hypertext capabilities
\usepackage[usenames,dvipsnames]{xcolor}
\usepackage{hyperref}
\hypersetup{
    colorlinks = true,
    citecolor = {magenta},
    linkcolor = {blue},
    urlcolor = {blue}		
}
\usepackage{mathtools}

\def\bi#1{\hbox{\boldmath{$#1$}}}

\begin{document}

\preprint{FERMILAB-PUB-21-306-T}

\title{
Beware of Fake $\bi{\nu}$s: \\
The Effect of Massive Neutrinos on the Non-Linear Evolution of Cosmic Structure
}

\author{Adrian E.~Bayer}
\email{abayer@berkeley.edu}
\affiliation{
 Berkeley Center for Cosmological Physics, University of California,
Berkeley,  CA 94720, USA
}%
\affiliation{Department of Physics, University of California,
Berkeley,  CA 94720, USA
}%

\author{Arka Banerjee}
\email{arka@iiserpune.ac.in}
\affiliation{Department of Physics, Indian Institute of Science Education and Research,
Homi Bhabha Road, Pashan, Pune 411008, India}
 \affiliation{Fermi National Accelerator Laboratory, Cosmic Physics Center, Batavia, IL 60510, USA}

\author{Uro\v{s} Seljak}
\email{useljak@berkeley.edu}
\affiliation{
 Berkeley Center for Cosmological Physics, University of California,
Berkeley,  CA 94720, USA
}%
\affiliation{Department of Physics, University of California,
Berkeley,  CA 94720, USA
}%
\affiliation{Physics Division, Lawrence Berkeley National Laboratory,  1 Cyclotron Road, Berkeley, CA 94720, USA}

\date{\today}

\begin{abstract}
Massive neutrinos suppress the growth of cosmic structure on small, non-linear, scales.
It is thus often proposed that using statistics beyond the power spectrum can tighten constraints on the neutrino mass by extracting additional information from these non-linear scales. We study the information content regarding neutrino mass at the field level, quantifying how much of this information arises from the difference in non-linear evolution between a cosmology with 1 fluid (CDM) and 2 fluids (CDM + neutrinos). We do so by running 
two $N$-body simulations, one with and one without massive neutrinos; both with the same phases, and matching their linear power 
spectrum at a given, low, redshift. This effectively isolates the information encoded in the linear initial conditions from the non-linear cosmic evolution.
We demonstrate that for $k \lesssim 1\,h/{\rm Mpc}$, and for a single redshift, there is negligible difference in the real-space CDM field between the two simulations. This suggests that all the information regarding  neutrino mass is in the linear power spectrum set by the initial conditions. Thus 
any probe based on the CDM field alone will have negligible constraining power beyond that which exists at the linear level over the same range of scales. Consequently, any probe based on the halo field will contain little information beyond the linear power.
We find similar results for the matter field responsible for weak lensing.
We also demonstrate that there may be much information beyond the power spectrum in the 3d matter field, however, this is not observable in modern surveys via dark matter halos or weak 
lensing.
Finally, we show that there is additional information to be found in redshift space.
\end{abstract}

\maketitle

%\tableofcontents

\section{Introduction}
\label{sec:intro}

Upcoming cosmological missions such as DESI\footnote{\url{https://www.desi.lbl.gov}}, Euclid\footnote{\url{https://www.euclid-ec.org}}, LSST\footnote{\url{https://www.lsst.org}}, PFS\footnote{\url{https://pfs.ipmu.jp/index.html}}, SKA\footnote{\url{https://www.skatelescope.org}}, and WFIRST\footnote{\url{https://wfirst.gsfc.nasa.gov/index.html}}, will probe progressively smaller scales of cosmic structure. It is hoped that by probing these small, non-linear, scales one will be able to detect much information regarding the total neutrino mass.
To fully realize the potential of these surveys, an urgent task is thus to quantify and optimally extract this information from the observed cosmological fields. 

%Instead of considering individual summary statistics that are computed from the underlying cosmological field, here we explore how much information there is regarding neutrino masses at the field level using more general arguments. We take a particular interest in fields that are experimentally observable in upcoming surveys, in order to motivate the focus of attention to the most promising avenues.

In a cosmology with massive neutrinos \cite{LESGOURGUES_2006}, we can define $\rho_{cb}$ as the contribution to the energy density due to cold dark matter (CDM) and baryons, $\rho_\nu$ as the contribution due to neutrinos, and $\rho_m$ as the total matter contribution. Given the lower bound on the sum of the neutrino masses coming from oscillation experiments is $M_\nu = 60 {\rm meV}$ \cite{SuperK, SNO, KamLAND, K2K, DayaBay}, neutrinos are non-relativistic at low redshift.
Defining $\bar{\rho}_X$ as the mean energy density in species $X$, where $X=\{cb,\nu,m\}$, we can further define the relative overdensity of species $X$ at redshift 0 as $\delta_X=(\rho_X-\bar{\rho}_X)/\bar{\rho}_X$, and the fraction of the total matter density in species $X$ as $f_X=\bar{\rho}_X/\bar{\rho}_m=\Omega_X/\Omega_m$.
This gives 
\begin{equation}
    \Omega_m\delta_m=\Omega_{cb}\delta_{cb}+\Omega_\nu\delta_\nu,
\end{equation}
 and
 the matter overdensity as
\begin{equation}
    \delta_m = (1-f_\nu) \delta_{cb} + f_\nu \delta_\nu.
    \label{eqn:delta_m}
\end{equation}

In practice, we cannot measure $\delta_\nu$ directly, as we 
do not have direct access to fluctuations in the cosmic neutrino background. We also cannot measure 
$f_\nu=\Omega_\nu/\Omega_m$ directly at low redshifts from the redshift-distance relations, since neutrinos 
are non-relativistic and their density has the same redshift dependence 
as cold dark matter and baryons. This leaves density 
perturbations in the total matter, $\delta_m$, and CDM+baryon, $\delta_{cb}$, fields as ways to probe neutrino mass
at low redshifts. 
So the success of upcoming surveys measuring neutrino mass hinges on their ability to measure the effects of neutrinos on the total matter and CDM+baryon perturbations, as well as on their ability to 
measure $\Omega_m$ from the redshift-distance 
relation (which can also be extracted 
from perturbations, such as from Baryonic Acoustic
Oscillations). 

On large scales neutrinos cluster analogously to CDM, whereas on small scales they do not cluster. The scale at which this transition occurs is known as the free streaming scale and is due to the neutrino thermal velocities erasing their own perturbations.
We can thus divide perturbations into scales 
larger than the neutrino free streaming scale, where $\delta_\nu \sim \delta_{cb}$, and 
scales smaller than that, where $\delta_\nu \sim 0$. 
One can see that 
if one could measure $\delta_m$ and $\delta_{cb}$ on small scales in the absence of noise, then any difference between the two would give strong constraints on neutrino mass via $\delta_m \sim (1-f_\nu)\delta_{cb}$. 
However, this poses several observational difficulties. 

A first difficulty is
that  the matter overdensity field $\delta_m$ is not directly observable.  
Weak lensing probes the
convergence, given by
\begin{align}
   % \nabla^2 \phi (\bi{x},z) = \frac{3}{2} a^2 H^2 (z) \Omega_m (z) \delta_m (\bi{x},z),
   \kappa(\chi_*,\hat{\bi{n}})
   &=\frac{3 H_0^2\Omega_{m}}{2c^2} \int_0^{\chi_*} d\chi ~
   \frac {\chi}{a(\chi)}\left(1-\frac{\chi}{\chi_*}\right)\delta_m(\chi \hat{\bi{n}}),
   %\frac{f_K(\chi_*-\chi)}{f_K(\chi_*)f_K(\chi)}  \delta_m (\chi \hat{\bi{n}}; \eta_0 - \chi),
   \label{eqn:kappa}
   %Omegam is Omegam0
\end{align}
where $\chi$ is the comoving distance, $\chi_*$ is the comoving distance to the source, $\hat{\bi{n}}$ is the direction on the sky, $H_0$ is the Hubble constant, $c$ the 
speed of light, $a(\chi)$ is the expansion factor, and we assume zero curvature.
Hence, $\kappa$ can be viewed as a measurement of $\Omega_m \delta_m$ averaged over a radial window along the line of sight between the observer and the source. This dilutes the information contained in the total matter field.

A second issue is that we also cannot measure 
$\delta_{cb}$ directly. What we can typically
measure from galaxy observations is a biased version, where at the linear level we have
$\delta_g=b_1\delta_{cb}$, with galaxy overdensity
$\delta_g$ being modulated by the linear bias $b_1$. The linear bias is constant on large scales, but has
complicated scale dependence on small scales which cannot be predicted ab initio and thus
has to be marginalized over to obtain constraints on cosmological parameters. 
One way to measure it is using redshift-space distortions (RSD), which at the linear order probes density-velocity correlations. Velocity can be related 
to the matter overdensity via $\delta_v=f\delta_m$, where $f$
is the linear growth rate which depends on 
the matter density $\Omega_m$. The growth rate is also affected by 
neutrinos, which slow down the growth 
of structure on small scales. However, on small scales, i.e. beyond linear order, this 
relation also becomes more complicated due to higher
order velocity-density correlators \citep[see e.g.][]{Chen_2021}, once again making it difficult to isolate the effects of neutrino mass. 

Multi-tracer analyses, combining $\delta_g$
from spectroscopic or photometric surveys, 
with weak lensing $\kappa$ from the cosmic microwave background (CMB) or large-scale structure (LSS), suggest that LSS surveys have 
the power to separate neutrino mass from 
other parameters, and that sampling variance 
cancellation is helpful on large scales \cite{Schmittfull_2018,yu2018neutrino}. Nevertheless, this approach is  
limited to about 20meV precision on the 
sum of neutrino masses for surveys such as 
LSST, suggesting it may not be able to 
give a neutrino mass detection at more than 
3 sigma for the minimum theoretical mass of 60meV. 

This limited precision from multi-tracer probes has revived interest
in measuring neutrinos from a single 
tracer using non-linear information. 
By studying the non-linear effects of massive neutrinos on structure formation \citep[][]{Saito_2008, Brandbyge_2009, Brandbyge_2010, shoji2010massive, Viel_2010, Ali_Ha_moud_2012, Bird_2012, hybrid, Costanzi_2013, Villaescusa_Navarro_2014, Villaescusa_Navarro_2018, Castorina_2014, Castorina_2015, Arka_2016, Archidiacono_2016, Carbone_2016, Upadhye_2016, Adamek_2017, Emberson_2017, Inman_2017, senatore2017effective, Yu_2017, Arka_2018, Liu2018MassiveNuS:Simulations, Dakin_2019, chen2020line, chen2020cosmic, Bayer_2021_fastpm},
several such statistics have 
been proposed, including the bispectrum, halo mass function, void size function, probability distribution function, and marked power spectrum \cite{Kreisch2019,liu&madhavacheril2019, Li2019, Coulton2019,Marques2019,ajani2020, Hahn_2020, hahn2020constraining, Uhlemann_2020, Massara_2020, bayer2021detecting, Kreisch_2021}. 
The reasoning is that a single tracer may 
have access to different types of information 
in different density regions. For example, while 
high density regions may be mostly sensitive 
to the CDM+baryons, which cluster and gravitationally collapse into virialized objects, low density regions 
such as voids may be more sensitive to 
neutrinos, which cluster weakly in comparison. Crucially, this implies that a full description of 
the system requires a two-fluid model, that of
CDM+baryons and of neutrinos, which cannot be 
mimicked by a single CDM+baryon component. 
The hope of this approach is that by effectively combining  information 
from different density regimes one might 
be able to determine neutrino mass to a much 
higher precision than predicted by just the 
two point statistics, the
power spectrum. 

The goal of this paper is to investigate this single tracer proposal
by comparing a single-fluid CDM simulation 
%whose
%initial setup is 
%as close as possible 
%matched
to a 
two-fluid simulation with CDM and neutrinos
(for the purpose of this paper we assume 
baryons trace CDM). We examine whether the presence of massive neutrinos has a unique \textit{non-linear} effect that differentiates the two at late times, or if the impact of the massive neutrino component can be faked by a solitary CDM component. To this end, we set up the
two simulations with matched
linear power spectrum of the field in question, and equal phases, %of CDM components 
at a redshift of interest, which 
we will take to be $z=0$. We 
compare the two simulations
at the field level for three different fields: (i) $\delta_{cb}$, which uniquely defines anything 
observable with galaxies, (ii) $\Omega_m \delta_m$, which is the corresponding field controlling 
weak lensing observables, and (iii) $\delta_m$, the 3d total matter field which is not currently observable.

If at the field level the two simulations 
differ in their phases 
%in the two 
%observables 
at $z=0$, this would suggest 
there is information that has been 
created by the non-linear evolution that is 
unique to the presence of massive neutrinos, and 
that cannot be mimicked by a single CDM fluid. 
If, on the other hand, the final phases are matched
exactly, 
then there is no information associated with the difference in non-linear evolution beyond the overall amplitude of the field, i.e.~the power spectrum. If the power spectra at $z=0$ are also identical between the two simulations then there is no non-linear information arising specifically from the presence of the neutrino component, and any information regarding neutrino mass must simply arise from the differing linear physics.
A similar analysis was performed in the context of modified gravity by \cite{Cataneo_2019i}, which studied only the non-linear power spectrum. Earlier work in the context of neutrino mass includes a study of the halo mass function  \cite{Cataneo_2019iii}, and the non-linear matter power spectrum for the Ly-$\alpha$ forest \cite{Pedersen_2020}. We will generalize such analyses by considering the information at the field level.

The structure of this paper is as follows. In Section \ref{sec:info} we outline how to study the information content of cosmological fields. In Section \ref{sec:neutrino} we apply this to understand the amount of neutrino mass information in the various aforementioned cosmological fields. In Section \ref{sec:otherobs} we then comment on the benefits of probes beyond the power spectrum (for example related to halos and voids). In Section \ref{sec:fisher} we consider a Fisher analysis to compare constraints obtained from the linear and non-linear power spectrum. Finally, in Section \ref{sec:conclusion} we conclude and discuss how our findings relate to constraints on $M_\nu$ presented in recent works.

\section{Cosmological Information}
\label{sec:info}

The simplest tool used to quantify the information content of a field $\delta(\bi{k})$ is the (auto) power spectrum $P_{\delta\delta}(k)$, defined via
\begin{equation}
    \langle \delta^*(\bi{k}) \delta (\bi{k}') \rangle = (2 \pi)^3 P_{\delta\delta}(k) \delta^{(D)} ( \bi{k} - \bi{k}'),
\end{equation}
where $\delta^{(D)}$ is the Dirac delta function.
$P_{\delta\delta}(k)$ is the Fourier transform of the two-point correlation function $\xi (r)$, i.e.~it measures the overdensity correlation between two arbitrary points of space separated by $r$. For a statistically homogeneous, isotropic, and Gaussian field, the power spectrum contains the entire information of the field. 
The standard model of cosmology assumes homogeneity and isotropy, and that the primordial Universe was described by a Gaussian random field (although we note that there are some extensions beyond this theory, for example positing primordial non-Gaussianity \citep{Maldacena_2003,Creminelli_2006,Komatsu_2009,Seljak_NG,Planck18_NG,Chen_2010,meerburg2019primordial}). 
The overdensity field in Fourier space is in general complex, i.e.~it can be written as $\delta(\bi{k})=|\delta(\bi{k})| e^{i\phi(\bi{k})}$, where $|\delta(\bi{k})|$ is the magnitude and $\phi(\bi{k})$ the phase. The phases of a Gaussian random field have a uniform random distribution in the range $[0,2\pi)$. 

The Universe then evolves, and during the late stages of evolution, structure formation introduces non-Gaussianities on small scales due to the non-linear nature of gravitational collapse.
The exact nature of this non-linear evolution depends on the cosmological parameters, for example the energy density of dark energy $\Omega_\Lambda$, the Hubble constant $H_0$, and the total neutrino mass $M_\nu$. There is thus much interest in studying higher-order statistics, in the hope that they contain additional information beyond the power spectrum. This is particularly true in the case of neutrinos due their signature on small, non-linear, scales. It is thus important to understand how much information neutrinos imprint on different cosmological fields, and furthermore how much of this information arises from non-linear cosmic evolution.

To set up the problem, let us consider two different universes at some late redshift $z_f$. We denote some generic field as $\delta_X(\bi{k},z_f)$ in the first universe with cosmological parameters $\lambda$, and $\tilde{\delta}_X(\bi{k},z_f)$ in the second universe with cosmological parameters $\tilde{\lambda}$.
A question of interest is, if our Universe corresponds to $\delta$, how well can we distinguish it from a universe with field $\tilde{\delta}$? Or in other words, how much information can we learn about the cosmological parameters by studying $\delta(\bi{k},z_f)$?
While a typical analysis, e.g.~a Fisher analysis, considers both linear and non-linear information as one, we seek to isolate the non-linear information.
More concretely, while a cosmological field may be sensitive to a change in cosmological parameters, if this sensitivity is purely at the linear level, then there will be no additional information compared to the linear power spectrum; one could consider non-linear probes, such as the halo mass function, void size function, the bispectrum, etc., but they will just be expressing the information content of the linear power spectrum in a different form.
So it is interesting to study how much non-linear information there is and thus how much benefit one can expect to extract from non-linear observables. 

To quantify how much non-linear information an entire field contains with regard to a change in cosmological parameters $\lambda-\tilde{\lambda}$, we match the linear physics at $z_f$ between the two cosmologies.
We then backscale the fields to some earlier redshift, $z_i$, using linear theory twice: one time using the cosmology associated with $\lambda$, and one time using $\tilde{\lambda}$. Finally we perform an $N$-body simulation to evolve the two fields to $z_f$ and obtain non-linear results: here again we use the appropriate choice of cosmology in each case.
A schematic of these two simulations is as follows:
\begin{align}
    %&\delta_{\mathcal{R}} \xrightarrow[Boltzmann]{\lambda} 
    &\delta^{(1)}_X(z_f) \xhookrightarrow[backscale]{\lambda} \delta^{(1)}_X(z_i) \xRightarrow[N-body]{\lambda} \delta_X(z_f), \label{Pevo} \\
    %&\delta_{\mathcal{R}} \xrightarrow[Boltzmann]{\lambda}
    &\delta^{(1)}_X(z_f) %\textcolor{blue}{
    \xhookrightarrow[backscale]{\tilde{\lambda}} \tilde{\delta}^{(1)}_X(z_i) \xRightarrow[N-body]{\tilde{\lambda}} \tilde{\delta}_X(z_f)
    %} 
    \label{Pevot},
\end{align}
where $\delta^{(1)}_X$ labels the linear power spectrum of component $X$.
The key difference between this approach and a typical analysis is the use of identical initial conditions for both universes to ensure the linear physics is the same at $z_f$ after running the simulation.
This means that any difference between $\delta_X(z_f)$ and $\tilde{\delta}_X(z_f)$ after the $N$-body simulation will be purely due to non-linear effects caused by using $\tilde{\lambda}$ instead of $\lambda$. 

Having set up the problem, we now review how to quantify the difference between two fields. Rather than considering specific observables, we seek to study effects at the field-level.
In order to compare the the two fields at $z_f$, we consider the (complex) coherence of the two fields, defined as 
\begin{equation}
    \zeta(k) = \frac{P_{\delta\tilde{\delta}}(k)}{\sqrt{P_{\delta\delta}(k)P_{\tilde{\delta}\tilde{\delta}}(k)}},
\end{equation}
where $P_{\delta\tilde{\delta}}(k)$ is the cross-power spectrum between $\delta$ and $\tilde{\delta}$, given by
\begin{equation}
    \langle \delta^*(\bi{k}) \tilde{\delta} (\bi{k}') \rangle = (2 \pi)^3 P_{\delta\tilde{\delta}}(k) \delta^{(D)} ( \bi{k} - \bi{k}').
\end{equation}
Unlike the auto power spectrum, the cross power spectrum can in general be complex.
Note that statistical isotropy and homogeneity enforces the coherence to only be a function of the magnitude $k$.

Two fields are said to be \textit{coherent} at scale $k$ if $|\zeta(k)|=1$. 
%In such a case the two fields are related by a linear transfer function, $T(k)$, given by
%\begin{equation}
%    T(k) = P_{{\delta}\tilde{\delta}}(k) / P_{\delta\delta}(k),
%    \label{P_transfer}
%\end{equation}
%such that $\tilde{\delta}(\bi{k},z) = T(k) \delta(\bi{k},z)$, 
%and consequently,
%\begin{equation}
%    P_{\tilde{\delta}\tilde{\delta}}(k) = |T(k)|^2 P_{\delta\delta}(k).
%    \label{P_transfer2}
%\end{equation}
In such a case the power spectra of the two fields are linearly related as follows
\begin{equation}
    P_{\tilde{\delta}\tilde{\delta}}(k) = \left|\frac{P_{{\delta}\tilde{\delta}}(k)}{P_{\delta\delta}(k)}\right|^2 P_{\delta\delta}(k),
    \label{P_transfer2}
\end{equation}
where the $|\cdot|^2$ term can be thought of as a linear transfer function between the auto spectra of the two fields.

If the real part of the coherence is equal to 1, the phases of $\delta$ and $\tilde{\delta}$ are statistically identical. If the phases of the two cosmologies evolved identically, then the entire difference between the two fields is captured by any difference in the amplitude of the individual power spectra.
Furthermore, if two fields are coherent, and the transfer function is identical to unity,
$|P_{{\delta}\tilde{\delta}}(k)/P_{\delta\delta}(k)|=1$, this implies that the power spectra are identical and that there is thus no non-linear information in the power spectrum. In such a case the two cosmologies are statistically indistinguishable in terms of non-linear effects, and there will be no information beyond the linear power spectrum. By this we mean that for a given set of scales, the information content of any non-linear statistic cannot exceed the information content of the linear power spectrum over those same scales. While the linear power spectrum is not something one can generally observe for a particular field, it is useful to know whether or not there is information that exists beyond linear theory.

\section{Massive Neutrino Information}
\label{sec:neutrino}

Using the notation of the previous section we use $\lambda$ to denote a universe with massive neutrinos, $M_\nu=0.15{\rm eV}$, and $\tilde{\lambda}$ to denote a universe without massive neutrinos, $M_\nu=0$. We start by using a Boltzmann solver to compute the linear power spectrum for a cosmology with $M_\nu=0.15{\rm eV}$ at $z_f=0$. 
We then backscale this power spectrum to $z_i=99$ twice, one time using the linear physics associated with massive neutrinos (using the REPS package \cite{Zennaro_2017}), giving $P_X(k,z_i)$, and the other time using the linear physics associated with massless neutrinos, giving $\tilde{P}_X(k,z_i)$. We generate realizations of the two fields at $z=99$ with matched phases. Note that the massless neutrino cosmology is thus initialized with a power spectrum whose shape encodes the linear suppression of growth due to the presence of massive neutrinos in the other cosmology. Then we evolve $P_X(k,z_i)$ through to $z_f$ using the Gadget $N$-body simulation \cite{Gadget} with massive neutrinos, yielding  $\delta_X(\bi{k},z_f)$, and we similarly evolve $\tilde{P}_X(k,z_i)$ through to $z_f$ using the $N$-body simulation without massive neutrinos, yielding  $\tilde{\delta}(k,z_f)$. Since the linear predictions of the two cosmologies have been matched as closely as possible, we can determine how much non-linear evolution is special to the presence of massive neutrinos by comparing the fields at $z_f$: ${\delta}(\bi{k},z_f)$ and $\tilde{\delta}(\bi{k},z_f)$. Furthermore, by comparing the power spectra of the fields, $P_{\delta\delta}(k,z_f)$ and $P_{\tilde{\delta}\tilde{\delta}}(k,z_f)$, we can assess the information in the power spectrum.
We refer to the $M_\nu > 0$ simulation as the ``real'' simulation, and the $M_\nu = 0$ simulation as the ``fake'' simulation, because the purpose of the $M_\nu = 0$ simulation is to fake the effects of massive neutrinos by using a single-fluid CDM simulation with initial conditions associated with a massive neutrino cosmology. Note that for each considered field, $\delta_X$,  a different fake $N$-body simulation is run with matched linear physics for that particular field.
We consider a box of volume $1\,({\rm Gpc}/h)^3$ and a grid of dimension $1024^3$ for both CDM and neutrinos.

In the case of lensing, the field $\delta_m$ is not directly measured. Instead, lensing measures $\Omega_m \delta_m$ averaged over a window function integrated over the line of sight, as described in Eqn.~(\ref{eqn:kappa}). We are therefore free to define the effective lensing field by rescaling by a constant factor, which we choose to be $(1-f_\nu)$ as follows:
\begin{equation}
    \kappa \sim \Omega_m \delta_m = \Omega_m(1-f_\nu)\frac{\delta_m}{{(1-f_\nu)}} = \Omega_c \frac{\delta_m}{{(1-f_\nu)}}.
\end{equation}
Assuming no a priori information regarding $\Omega_m$, we can evaluate the lensing information by considering the information in the field defined by
\begin{equation}
    \delta_{\Omega m} \equiv \frac{\delta_m}{(1-f_\nu)}
    \label{eqn:delta_Om}.
\end{equation}
While we are free to choose any normalization, the reason for this choice is that we seek the option which most closely matches the real and fake cosmologies. From Eqn.~(\ref{eqn:delta_m}) it is clear that $\delta_{\Omega m} \sim \delta_{cb}$ on small scales, thus this choice is inspired such that neutrino effects should be negligible on small scales (note that in the case of $\delta_{\Omega m}$ linear matching cannot be achieved at large scales, but rather on small scales).% at the linear level. %Any difference observed in the non-linear field will thus be due to nonlinear effects.

\begin{figure}[t]
\includegraphics[width=\linewidth]{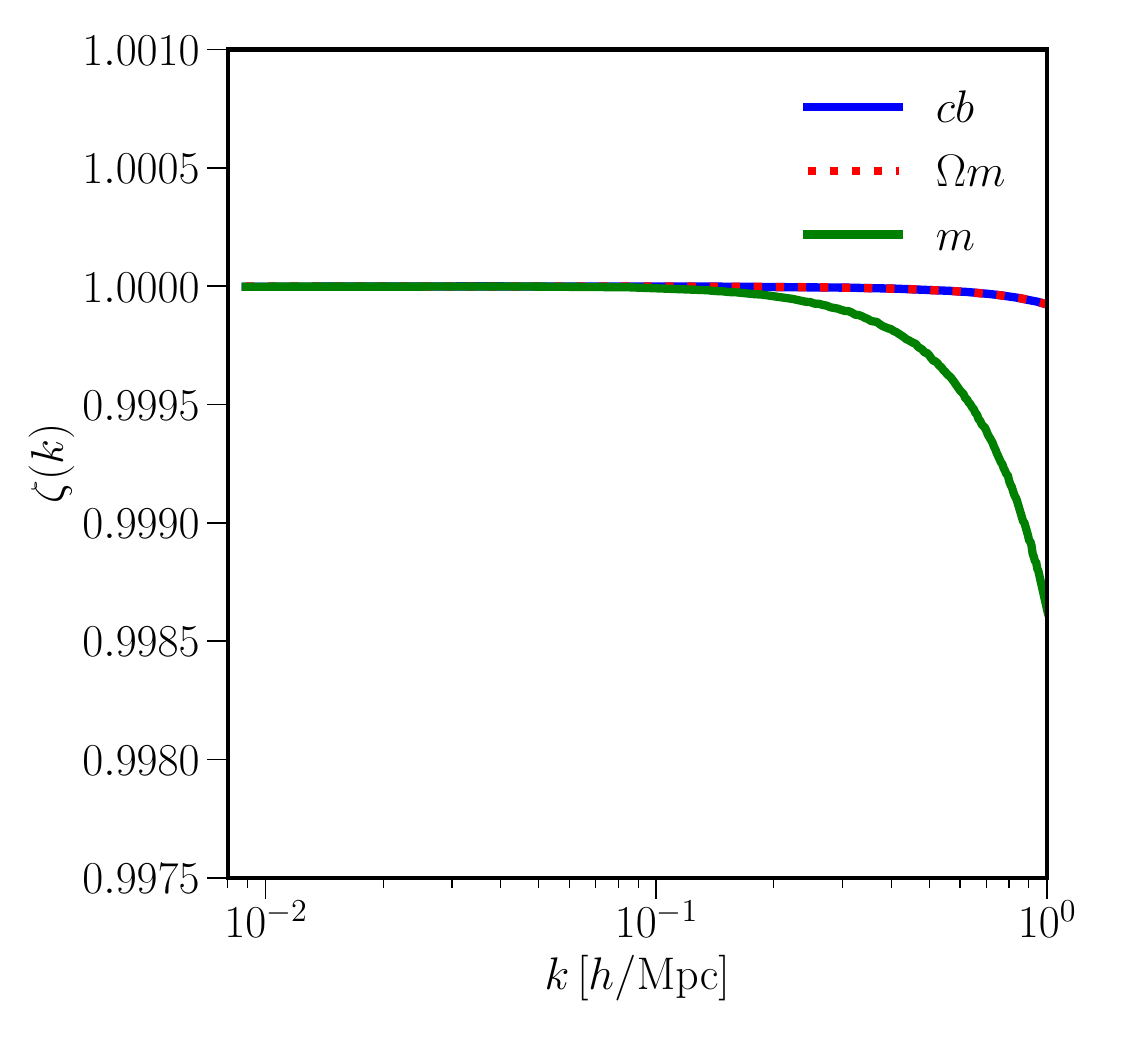}
\caption{The real part of the coherence between fields from the real and fake simulations. If we match $\delta_X$ at $z=0$, we plot the coherence for $\delta_X$. It can be seen that the coherence in the case of $\delta_{cb}$ and $\delta_{\Omega m}$ is 1 up to $k=1\,h/{\rm Mpc}$ to $\lesssim 0.01\%$. This implies that there is negligible non-linear information in the $cb$ field or the lensed matter field at these scales that goes beyond the power spectrum.
On the other hand, the coherence for $\delta_m$ begins to differ from 1 at a lower value of $k$, implying non-linear information beyond the power spectrum for the 3d matter field.
}
\label{fig:r_cc} 
\end{figure}

In Fig.~\ref{fig:r_cc} we plot the real part of the coherence for each of the fields between the real and fake simulations. Specifically, when we match $P_X$ at $z=0$, we plot the coherence for the $X$ overdensity field.
It can firstly be seen that in the case of $X=cb$, the coherence is unity up to $k=1\,h/{\rm Mpc}$ to $\lesssim 0.01\%$. 
This implies that the final phases of the $cb$ field are equivalent regardless of whether massive neutrinos are included in the simulation. This is to say that the non-linear evolution of the $cb$ field is identical in both the one-fluid (CDM) and two-fluid (CDM+$\nu$) description. Thus
%This implies that 
there is negligible non-linear information in the $cb$ field that goes beyond the power spectrum within scales of experimental interest.
%This means that all the phases are identical, not just $P_{cb}(k)$, implying that there is no non-linear information in the $cb$-field that goes beyond the power spectrum within at scales of experimental interest.
On the other hand, the coherence for $\delta_m$ begins to differ from 1 at a lower value of $k$, implying there is non-linear information beyond the power spectrum for the 3d matter field. 
%Such information has been found in works such as \cite{Uhlemann_2020, Massara_2020, bayer2021detecting}. 
However, one cannot measure the matter field directly, and one instead measures lensing which is related to $\delta_{\Omega m}$, as in Eqn.~(\ref{eqn:delta_Om}). In this case the picture is identical to $\delta_{cb}$, with a coherence of 1 up to $k=1\,h/{\rm Mpc}$ to $\lesssim 0.01\%$, implying negligible non-linear information beyond the power spectrum in this field at these scales.

\begin{figure}[t]
\includegraphics[width=\linewidth]{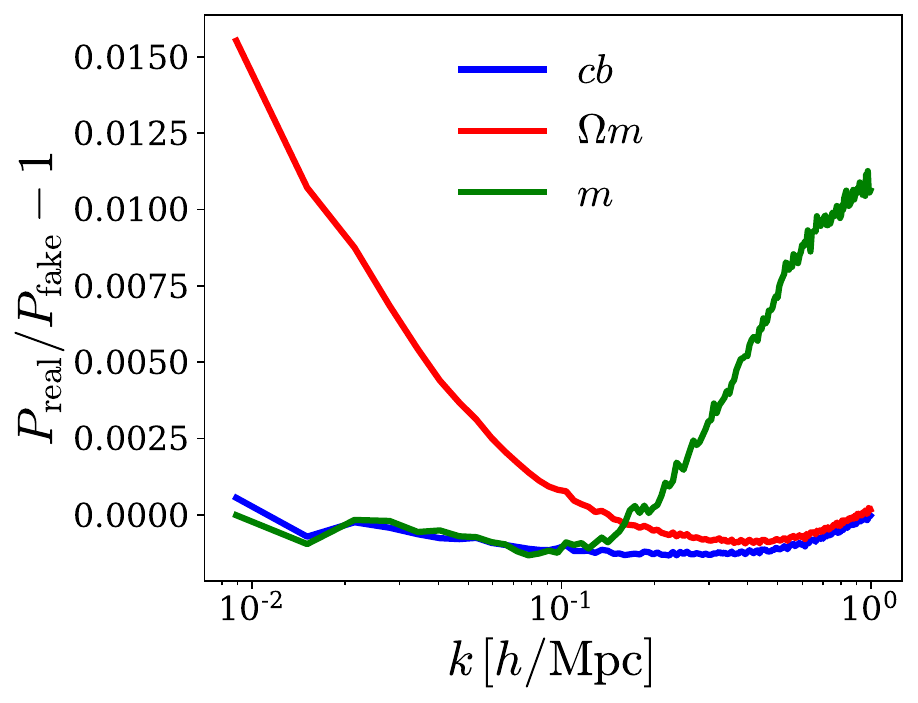}
\caption{The ratio of the power spectra between the real and fake simulations. If we match $\delta_X$ at $z=0$, we plot the corresponding power spectrum, $P_X$. It can be seen that the ratio is 1 for $cb$, while there is an approximately 1\% deviation for $P_{\Omega m}$ on large scales. Only $P_m$ differs form 1 on non-linear scales, implying information beyond the power spectrum in for the matter field. In the cases of $cb$ and $\Omega m$, the $\lesssim 0.1\%$ upturn on scales smaller than $k\approx0.5\,h/{\rm Mpc}$ is a numerical artifact due to discrepancy between the backscaling and forward model; a similar effect can be seen in the case of $m$ for which a downturn begins at $k\approx0.5\,h/{\rm Mpc}$.}
\label{fig:P_ratio} 
\end{figure}

Having established that, for scales of interest, the information in the case of $\delta_{cb}$ and $\delta_{\Omega m}$ is all in the power spectrum, we now consider how much information the power spectrum contains. In Fig.~\ref{fig:P_ratio} we plot the ratio between the power spectra for the various fields at redshift $z_f$. We see that for $P_{cb}$ the ratio is always 1, implying that there is negligible non-linear information regarding neutrinos in the $cb$ power spectrum. 
(We note that the $\lesssim 0.1\%$ upturn for scales smaller than $k\approx0.5\,h/{\rm Mpc}$ is a numerical artifact caused by a slight discrepancy between the growth factor implemented in backscaling and that effectively implemented by the $N$-body simulation. The magnitude of this discrepancy depends on $M_\nu$, leading to this small effect.)
Given the coherence of the $cb$ field is 1, this means there is negligible non-linear information about neutrino mass in the entire $cb$ field. On the other hand, there is a deviation of order 1\% in $P_{\Omega m}$ for $k \lesssim 0.1\,h/{\rm Mpc}$. This implies there is some information on neutrino mass in the lensed matter power spectrum. This is the typical shape information associated with neutrinos, however, it mostly appears on large, linear, scales and will thus be sample variance limited. Finally, we see that the ratio for $P_m$ differs from 1 on small scales, implying the presence of non-linear information about neutrinos beyond the the linear power spectrum of the 3d total matter field. 

To summarize, whenever we consider the single-fluid CDM field, we find that there is no difference between the real and fake simulations. On the other hand, whenever we consider fields that explicitly depend on both fluids (CDM+$\nu$) in the real simulations, we generally find that a single fake simulation cannot reproduce the statistics on all scales: either they remain matched on large scales, or they remain matched on small scales. The two choices we explore, $m$ and $\Omega m$, illustrate this clearly. For $m$, the small scales have a different non-linear behavior even though the linear statistics are exactly matched. For $\Omega m$, the large scales are not matched even at the linear level, but crucially, the small scale matching is maintained both at the linear and non-linear level.
Since for $\Omega m$ only the linear scales are not matched well, most of the information should be contained in the linear power spectrum.

\section{Higher-Order Statistics}
\label{sec:otherobs}

We now illustrate the effect of the results of the previous section on various statistics beyond the power spectrum. While the results of the previous section are sufficient in determining the presence of information regarding $M_\nu$ in any non-linear statistic beyond the power spectrum, we now show this explicitly for various examples in the interest of clarity.

\begin{figure}[t]
\includegraphics[width=\linewidth]{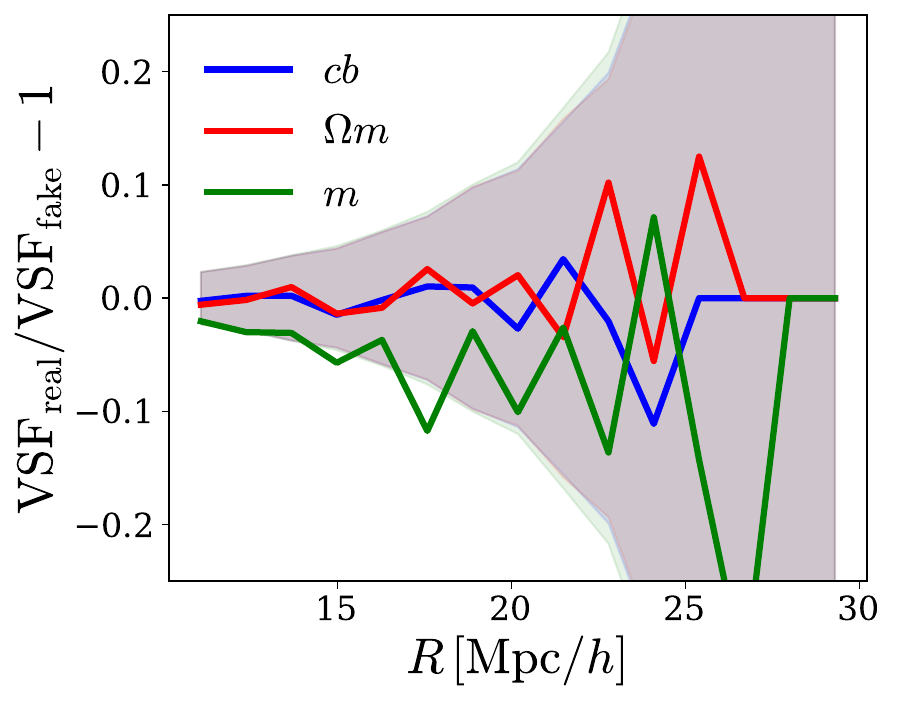}
\caption{The ratio of the void size function between the real and fake simulations. If we match $\delta_X$ at $z=0$, we plot the corresponding VSF in the $X$ field. Bands represent Poisson errors. It can be seen that the ratio is 1 for $cb$ and $\Omega m$ within the Poisson errors. Only the VSF in the 3d matter field shows a ratio that isn't unity, although it is still close to the Poisson error.}
\label{fig:V_ratio} 
\end{figure}

We start with the void size function (VSF), a commonly proposed source of information regarding neutrino mass \cite{Kreisch2019}. We use the spherical void finder of \cite{Arka_2016} with a threshold of $\delta_{\rm th}=-0.7$, and look for voids in the three considered fields: $cb$, $\Omega m$, and $m$. In Fig.~\ref{fig:V_ratio} we find that there is no difference in the VSF between the real and fake simulations for both the $cb$ and $\Omega m$ fields, but there is potentially some difference for the 3d matter field. Note that we find similar results regardless of the value of $\delta_{\rm th}$.

\begin{figure}[t]
\includegraphics[width=\linewidth]{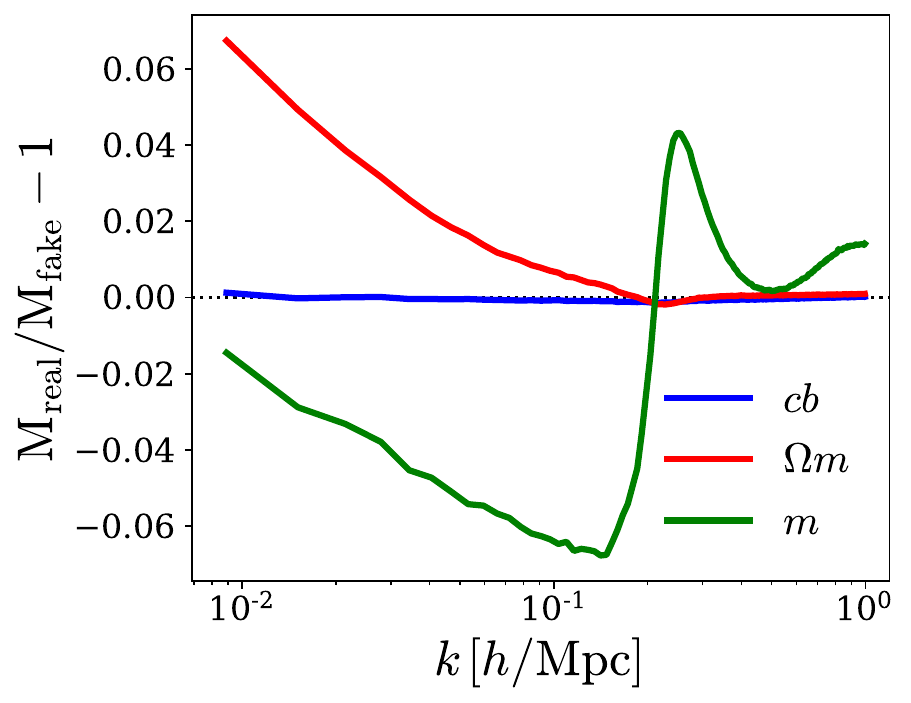}
\caption{The ratio of the marked power spectrum ($M$) between the real and fake simulations. If we match $\delta_X$ at $z=0$, we plot the corresponding $M$ in the $X$ field. It can be seen that the ratio is 1 for $cb$, while for the $\Omega_m$ it deviates form 1 on large scales, and for $m$ is deviates from 1 on all scales.}
\label{fig:M_ratio} 
\end{figure}

\begin{figure}[t!]
\includegraphics[width=\linewidth]{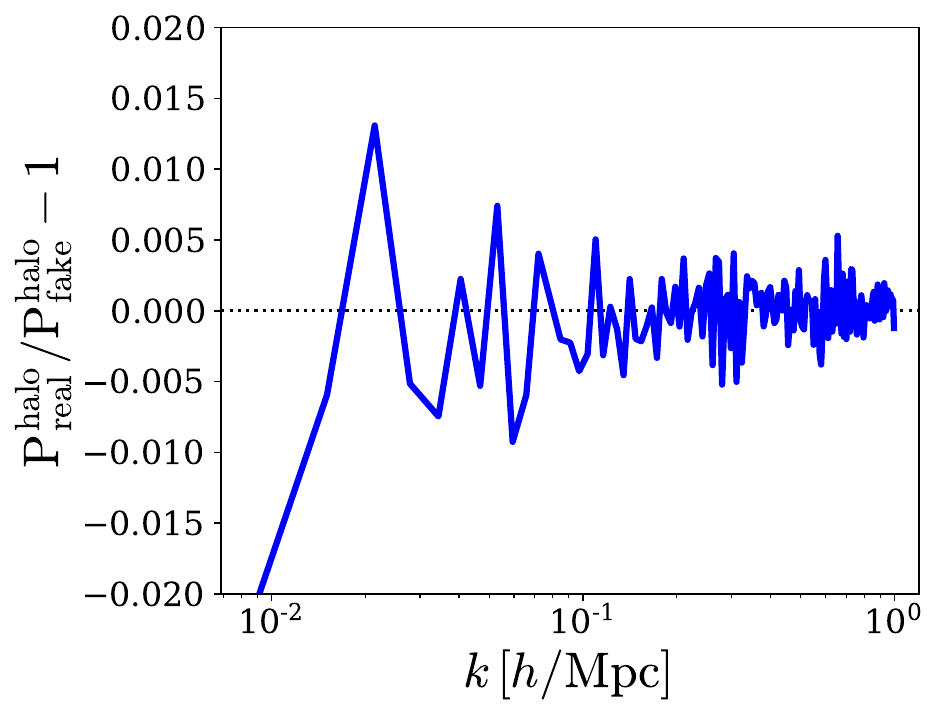}
\includegraphics[width=\linewidth]{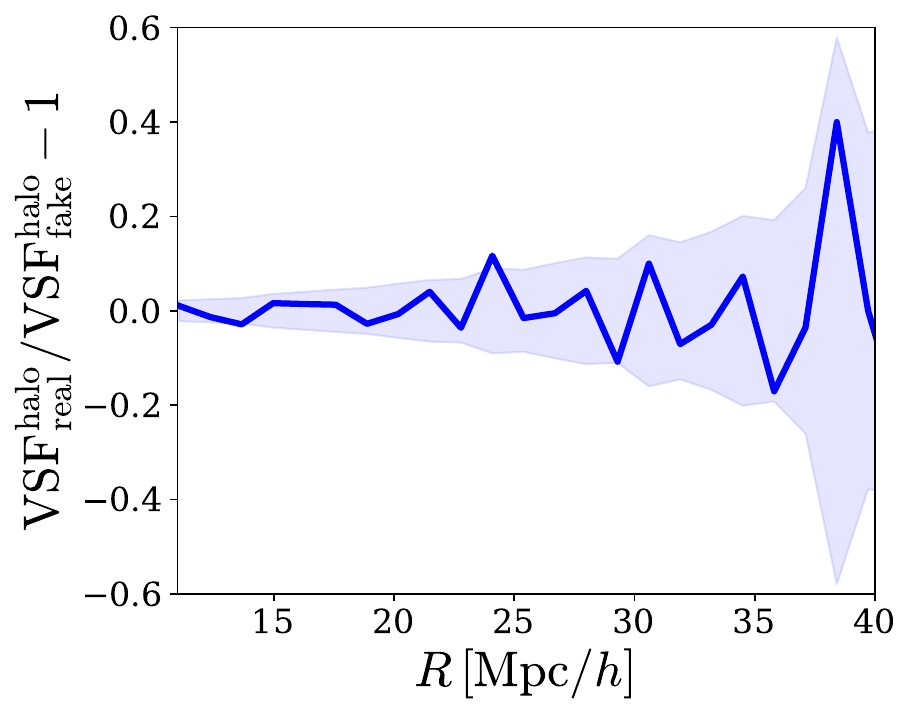}
\includegraphics[width=\linewidth]{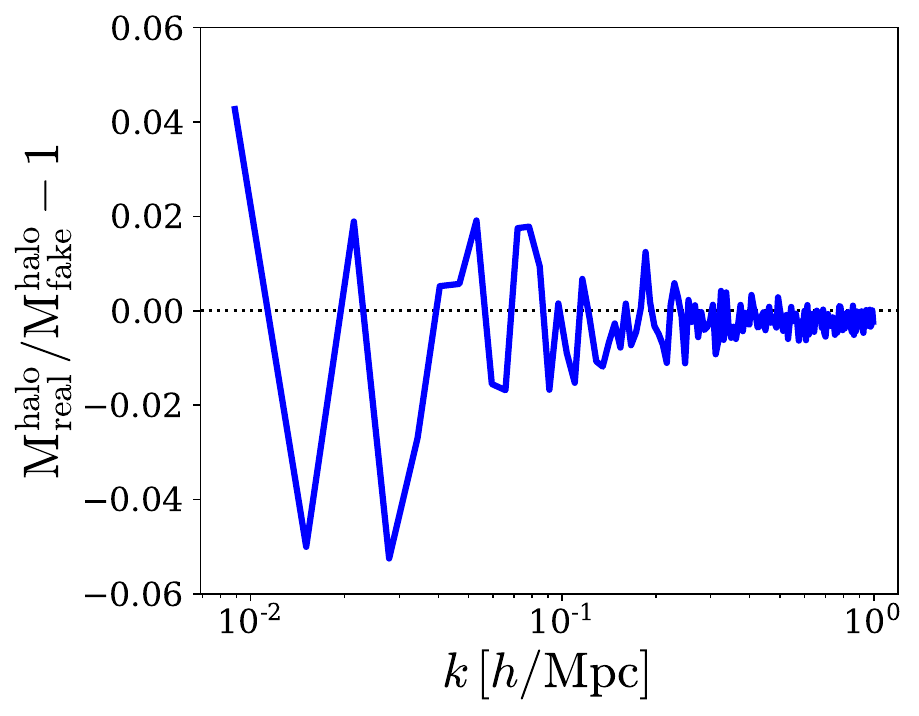}
\caption{The ratio of the halo-traced power spectrum, void size function, and marked power spectrum (from top to bottom) between the real and fake ($cb$-matched)  simulations. It can be seen that the ratio is close to unity in all cases.}
\label{fig:halos} 
\end{figure}

\begin{figure*}[t!]
\centering
\includegraphics[width=0.49\textwidth]{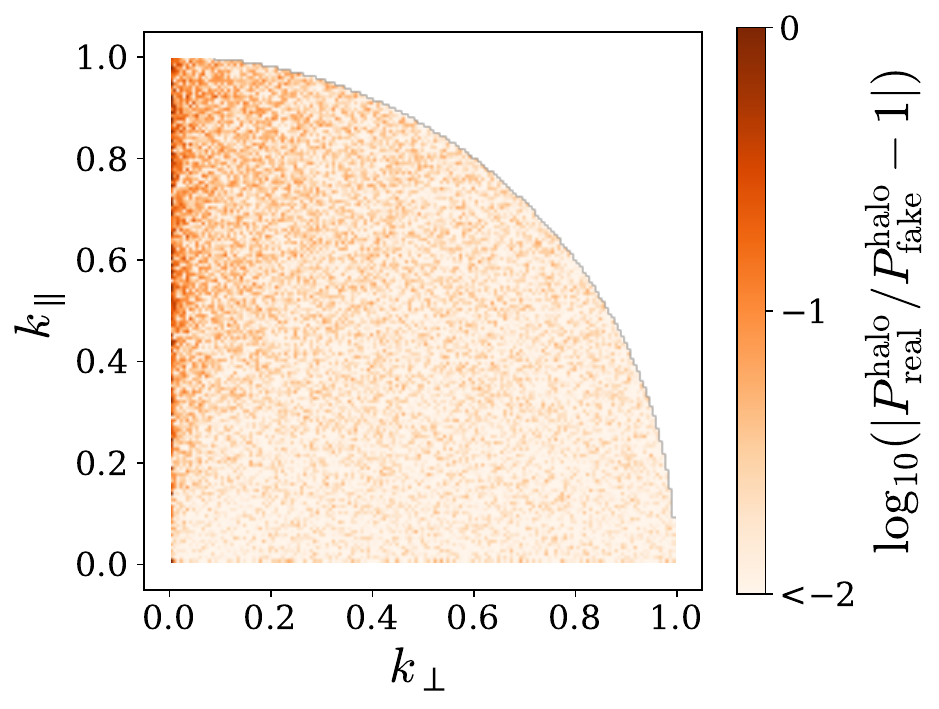}
\hspace{\fill}
\includegraphics[width=0.49\textwidth]{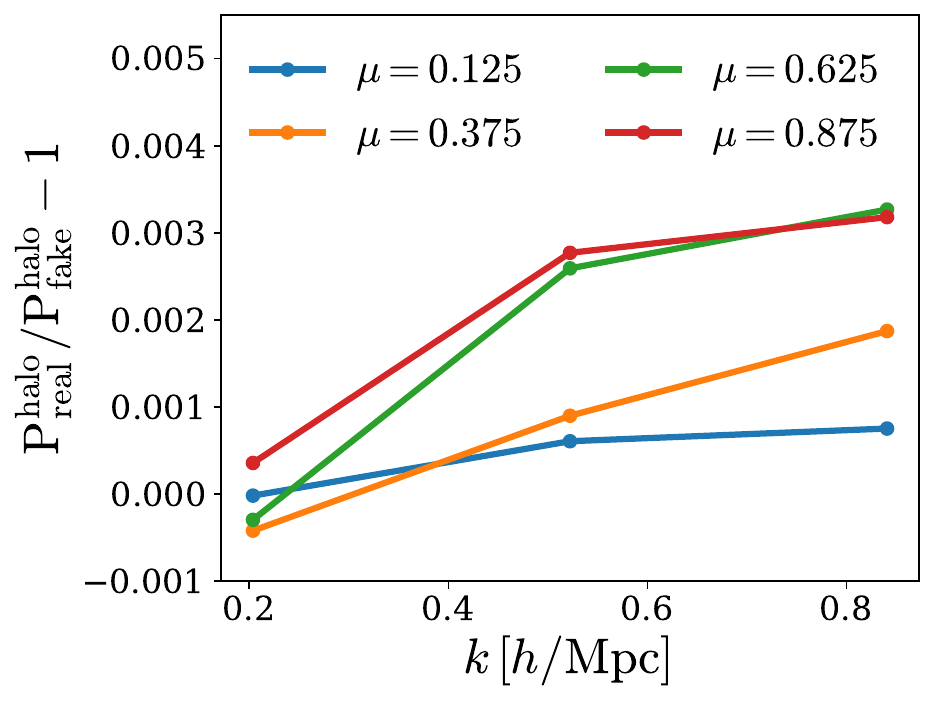}
\caption{The ratio of the redshift-space halo power spectrum between the real and fake ($cb$ matched)  simulations. \textit{Left}: bin-by-bin comparison in the $(k_\parallel,k_\perp)$ plane, where parallel/perpendicular is in reference to the LoS. \textit{Right:} binning the data into 3 bins of magnitude, $k$, and 4 bins of LoS projection, $\mu=k_\parallel/k$. Both plots show a deviation of the ratio from unity when moving closer to the LoS and to smaller scales, but negligible deviation in the perpendicular direction, suggesting that the additional information on neutrino mass comes from the modified velocity field, or growth rate, which is sourced by the matter overdensity.}
\label{fig:RSD} 
\end{figure*}

Next, in Fig.~\ref{fig:M_ratio}, we consider the marked power spectrum. We use the optimal choice of mark parameters found in \cite{Massara_2020}, which uses the smoothed overdensity field with $10 \,{\rm Mpc}/h$ smoothing window, thus small scale information is mixed into large scales. We again find little difference in the $cb$ field. The $\Omega m$ field differs only on large, linear, scales. For the $m$ field there is a difference on all scales. 
Again, this fits with our findings in the previous section.

A corollary of there being negligible information in the $cb$ field is that there will also be negligible information in the halo field. The halo field is a function of the $cb$ field and the bias parameters, hence, without knowledge of the bias, the information content of the halo field is just a re-expression of the information contained in the $cb$ field. We illustrate this in Fig.~\ref{fig:halos}, which shows the difference in the power spectrum, void size function, and marked power spectrum, for the halo field between the real and fake ($cb$-matched) simulations. We identify halos using the Friends-of-Friends (FoF) algorithm and apply a fixed number density cut. We find that there is no significant difference in any of the halo statistics between the real and fake simulations.

Having shown there to be little information in the real-space halo field, we now consider redshift-space distortions (RSD), which includes the effects of the peculiar velocities of halos along the line of sight (LoS). The peculiar velocity field is sourced by the clustering of the total matter field, which includes neutrinos. The halo power spectrum in redshift space can, therefore, provide additional information on the total neutrino mass.
%These introduce a line-of-sight (LoS) distortion due to the Doppler effect, and thus depends on the particle velocities, which in turn are determined by the clustered matter overdensity. Also, neutrinos have very different thermal velocity profiles compared to CDM. Thus one can use this information to better constrain the neutrino mass. 
In the left panel of Fig.~\ref{fig:RSD} we show a bin-by-bin comparison of the redshift-space halo power spectrum from the real and fake (with matched $cb$ field) simulations in the $(k_\parallel,k_\perp)$ plane, where parallel/perpendicular is in reference to the LoS. It can be seen that some bins along the LoS have a relatively large difference between the two simulations, but even small deviations from the LoS direction brings the size of the effect down to $\lesssim 1 \%$, in  line with the results obtained in real space. To better visualize the dependence on magnitude, $k$, and projection onto the LoS, $\mu=k_\parallel/k$, we bin the data into 3 bins of $k$ and 4 bins of $\mu$. The right panel shows the difference between the real and fake simulation increases with $k$ and $\mu$, signifying the information present at small scales due to RSD as one approaches the LoS. Therefore, we conclude that there is indeed additional non-linear information about neutrino mass that can be obtained by studying clustering of biased tracers in redshift space. While this clustering can be difficult to model accurately, it may be a key source of information in upcoming surveys \cite{DESInu}.

\section{Fisher Analysis}
\label{sec:fisher}

As shown the previous two sections, without RSD, the $cb$ field is statistically indistinguishable between the one-fluid (CDM) and two-fluid (CDM+$\nu$) simulations. The same is also approximately true for the $\Omega m$ field, responsible for weak lensing, for which there is only a difference in the power spectrum on large scales. If there is negligible difference between the one-fluid and the two-fluid non-linear dynamics, the 
total information content is essentially maximized by 
that which arises from the linear physics.
Nevertheless, as the linear power spectrum is not observable, it is instructive to compare the information content of the linear power spectrum to the non-linear power spectrum.

We perform a Fisher analysis
in the $\{\Omega_m, \Omega_b,h,n_s, \sigma_8, M_\nu\}$ plane. 
We use a fiducial cosmology with $\Omega_m = 0.3175$, 
$\Omega_b=0.049$, $h=0.6711$, $n_s=0.9624$, 
$\sigma_8=0.834$, and $M_\nu=0.05 {\rm eV}$. To compute derivatives we use a central difference scheme at $\pm \delta \theta$ for each cosmological parameter. Specifically we use $\delta \Omega_m = 0.01$, $\delta\Omega_b=0.002$, $\delta h=0.02$, $\delta n_s=0.02$, 
$\delta\sigma_8=0.015$, and $\delta M_\nu=0.025 {\rm eV}$. For the linear covariance between probes $x$ and $y$, we use $C_X = 2 P_X^2 / N_k$, where $N_k = 4 \pi  k^2 k_F / k_F^3$, and $k_F = 2 \pi / L$ is the fundamental wavenumber which we take for a box of volume $1 \, ({\rm Gpc}/h)^3$. 
For the non-linear results we use the Quijote simulations \citep{quijote}, and for the linear results we use CAMB \citep{CAMB}, using the same derivative computation method and binning as Quijote. 

\begin{figure}[!t]
\includegraphics[width=\linewidth]{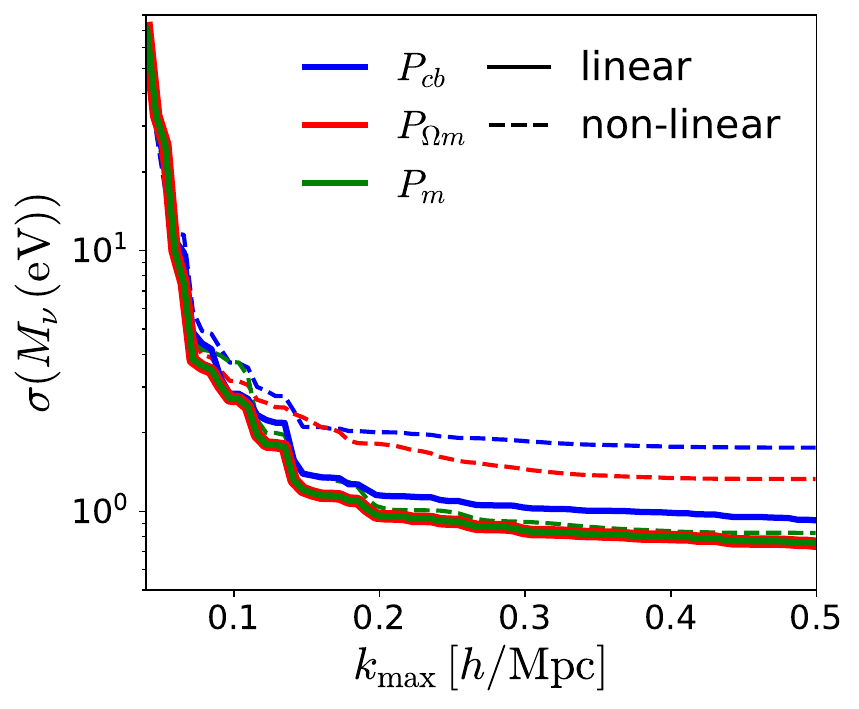}
\caption{Marginal error on $M_\nu$ for $P_{cb}$ (blue), $P_{\Omega m}$ (red), and $P_m$ (green), in both the linear (solid) and non-linear (dashed) regime, for a volume of $1\,({\rm Gpc}/h)^3$. }
\label{fig:Fisher} 
\end{figure}

Fig.~\ref{fig:Fisher} shows the marginal error on $M_\nu$ as a function of $k_{\rm max}$ for the various linear and non-linear power spectra. As expected there is good agreement between the linear and non-linear results on large scales, where cosmic evolution is approximately linear. Moving to smaller scales, we see that the non-linear power spectra for $cb$ and $\Omega m$ have a factor 2 times lower constraining power compared to their linear counterparts. Note that the non-linear power has worse constraints because its covariance has positive off-diagonal elements due to mode coupling, which in turn degrades the information content after marginalizing \cite{bayer2021detecting}. This means that there is still potential room for improvement upon the constraints from the non-linear power spectrum, and one may benefit from around a factor 2 by using statistics beyond the power spectrum. Thus, on their own, the $cb$ and $\Omega m$ fields give a marginal error on the neutrino mass of just under 1eV in a $1\,({\rm Gpc}/h)^3$ volume. 

On the other hand, the linear and non-linear marginal error on $P_m$ match well all the way to $k_{\rm max}=0.5\,h/{\rm Mpc}$. But, regardless of this, it was shown in the previous section that there is additional information in the phases of the 3d matter field that is not fully 
captured by the power spectrum, and there is thus additional information to be found in higher-order statistics.

\section{Discussion and Conclusions}
\label{sec:conclusion}

In this paper we have investigated how much non-linear information regarding neutrino mass one can expect to find in various cosmological fields by comparing one-fluid (CDM) to two-fluid (CDM+$\nu$) simulations with matched initial conditions. In real space, we found that the $cb$ field and $\Omega m$ (lensing) field %contains 
%no information beyond the linear power spectrum at the $\mathcal{O}(0.01 \%)$ level up to $k<1\,{\rm Mpc}/h$. %Additionally, the $cb$-field contains no information beyond the linear power spectrum, which 
do not contain additional information regarding neutrino mass that is unique to the 
two-fluid dynamics up to $k\lesssim 1\,h/{\rm Mpc}$. Essentially, the evolutionary effect of including a massive neutrino fluid can be faked by a solitary CDM fluid.
This implies that the $cb$ field, and derived quantities (e.g.~the halo field), and weak-lensing convergence,  contain little information regarding neutrino mass beyond that which exists in the linear power spectrum over the same scales. 
We have also shown that there is much non-linear information regarding neutrino mass in the 3d matter overdensity field,
%--- as found in \cite{Uhlemann_2020, Massara_2020,bayer2021detecting} --- 
however, this is not currently experimentally detectable.
The fundamental quantities we considered are the coherence and power spectrum ratio between the two simulations, summarized in Table \ref{tab:summary}, which alone quantify the amount of non-linear information at the field level. We then explicitly verified these findings for various higher-order statistics, including the void size function and marked power spectrum.

\begin{table}[b]
\caption{\label{tab:summary}
Summary of key results. The coherence and power spectrum ratio between the real and fake simulations for the $cb$, $\Omega m$, and $m$ fields, for $k \leq 1\,h/{\rm Mpc}$. Note that while the power spectrum ratio for the $\Omega m$ field differs from unity at the 1\% level, this is only at low $k$ which is sample variance limited.
}
%\begin{ruledtabular}
\begin{tabular}{l|cc}
\textrm{Field}&
\textrm{$\zeta-1$}&
\textrm{$P_{\rm real}/P_{\rm fake}-1$}\\
%\textrm{Right}\\
\colrule
$cb$ & $\lesssim 0.01 \%$ & $\lesssim 0.1 \%$ \\
$\Omega m$ & $\lesssim 0.01 \%$ & $\lesssim 1 \%$ (low $k$) \\
$m$ & $\lesssim 0.1 \%$ & $\lesssim 1 \%$ \\
\end{tabular}
%\end{ruledtabular}
\end{table}

%We have shown that the real-space galaxy field and weak-lensing convergence will not contain much information regarding $M_\nu$ on their own, and that this information content is effectively bounded by that which exists in the linear power spectrum.
Consequently, one can expect constraints on neutrino mass a little lower than 1eV in a volume of $1\,({\rm Gpc}/h)^3$ when using the $cb$ or lensing fields alone at a single redshift. Hence, using only this information, a very large volume of $10^4\,({\rm Gpc}/h)^3$ would be needed to 
reach an error of 0.01eV (corresponding to a $\sim 5$ sigma detection), which exceeds the available volume of currently realistic surveys. 

We note that even in the face of these findings, there is still motivation to consider statistics beyond the power spectrum to detect neutrino mass.
A first consideration is the choice of redshift(s). In our analysis we have matched the linear physics at a single redshift, $z=0$. Similarly, \cite{Pedersen_2020} used hydrodynamical simulations to find one can fake the effects of massive neutrinos in the non-linear power spectrum up to $k\lesssim10/{\rm Mpc}$ for the Ly-$\alpha$ forest ($z=3$). While the effects of massive neutrinos can be faked at a single redshift, the real and fake universes have, in principle, different evolution. Therefore, combining the fields at multiple redshifts should help discriminate between the two and improve constraints on $M_\nu$.
It will thus be imperative to combine multiple redshifts (from CMB redshift of 1100 to today)
and tracers (CMB, galaxies, and weak lensing) to obtain tight constrains on neutrino mass in upcoming surveys.
For example, combining
weak lensing and galaxy clustering, can reach 0.02eV with Rubin (LSST) and Stage IV CMB \cite{Schmittfull_2018}.

Second, the late-time linear power spectrum is not an observable quantity, as cosmic evolution is indeed non-linear. Hence, even though the linear power spectrum may provide a bound for the error in the case of $cb$ and lensing, the non-linear power spectrum does not quite reach this bound. We have shown this effect corresponds to around a factor of 2,
%Furthermore the covariance matrix of the non-linear power spectrum introduces positive correlations between modes, which often reduces the Fisher 
thus a different non-linear statistic may be able to obtain slightly better constraints than the non-linear power spectrum. This factor of 2 can also be recovered by reconstructing the linear field from the non-linear field \cite{Seljak_2017}.

Third, we consider lensing measurements to be directly sensitive to the product $\Omega_m \delta_m$. This is exact if the sources are 
at low redshift.
However, the comoving distance in Eqn.~(\ref{eqn:kappa}) implicitly depends on $\Omega_m$ as well, so for sources at higher 
redshift the relation is more complicated. Thus all possible combinations of $\Omega_m$ and $\delta_m$ that keep $\Omega_m \delta_m$ fixed may not be compatible with the observed lensing signal because they will modify the comoving distances. If one could obtain strong constrains on $\Omega_m$ from the redshift-distance relation, then combining it with lensing measurements may be able to probe $\delta_m$ directly, rather than the product $\Omega_m \delta_m$. 
We also note that neutrinos are non-relativistic at low redshift and thus will not induce a significant geometric effect on lensing observables which is known to arise in the context of dark energy \cite{Simpson_2005, Matilla_2017}.

Fourth, we have motivated that redshift-space distortions (RSD) may provide non-linear information regarding neutrino mass, thus considering higher-order statistics in redshift space is a worthwhile pursuit. RSD adds new information 
because velocities are determined by the growth factor $f$, which is 
sensitive to matter density $\Omega_m$ and neutrino density $\Omega_\nu$.
While RSD can be difficult to model, it could be a key source of information in upcoming surveys \cite{DESInu}. For example, \cite{kuruvilla2021information} illustrates how halo velocities can aid in constraining neutrino mass.
A further 
improvement on $f$ may be possible from redshift dependence, which 
we did not consider in this paper. 
%, or from the non-linear 
%evolution, which is controlled by $f$.

Fifth, it might be argued that even for the $cb$ or $\Omega m$ fields one could find information regarding $M_\nu$ beyond the linear power spectrum, as there may be a non-linear statistic with more favourable parameter degeneracies. For example, a particular non-linear statistic might constrain some other cosmological parameter much better than the linear power spectrum, and thus after marginalizing over this parameter, the constraints on $M_\nu$ will outperform the linear power spectrum.
However, the other parameters of key relevance in the case of neutrino mass are $\Omega_m$ and $A_s$, for which non-linear cosmic evolution does not induce additional information beyond that which exists in the linear initial conditions. We illustrate this in Appendix \ref{app:params}. Thus it is not expected that degeneracies will cause a big improvement in the constraints on $M_\nu$.
One could also consider non-standard cosmological parameters, for example related to primordial non-Gaussianty or exotic neutrino interactions \cite{Kreisch_2020}. For the latter to have an effect there would likely need to be a mechanism which couples the non-linear evolutions of the $cb$ and neutrino perturbations much more strongly than what happens through the Poisson equation. In principle this is possible given a sufficiently strong neutrino-baryon or neutrino-neutrino interaction, and this could help break-degeneracies with neutrino mass if one had a means to measure this non-standard effect.

Numerous recent works have proposed that one can obtain information regarding  neutrino mass beyond the power spectrum %in both the CDM+baryon and total matter fields 
\cite{liu&madhavacheril2019, Li2019, Coulton2019,Marques2019,ajani2020, 
Hahn_2020, hahn2020constraining, 
Uhlemann_2020, Massara_2020, bayer2021detecting, Kreisch_2021, Cheng_2021, Valogiannis_2021}. Some forecast $\mathcal{O}(0.1 {\rm eV})$ constraints by employing tomography, which is in good agreement with our results. On the other hand, some works find constraints that are over an order of magnitude smaller than linear theory. % $\sim$1eV in a $1\,({\rm Gpc}/h)^3$ volume with a single tracer. 
Given our findings we are able to explain exactly where this information comes from. 
In the case of \cite{Hahn_2020, hahn2020constraining} the information arises from working in redshift space, while for \cite{Uhlemann_2020, Massara_2020, bayer2021detecting} it comes from working with the 3d matter field.
Regarding \cite{Kreisch_2021}, which considers the real-space halo field, the information comes from assuming knowledge of the bias model as a function of cosmology.
The bias model can be thought to transfer information on small scales in the $cb$ field to larger scales in the halo field, thus information at $k > 1\,h/{\rm Mpc}$ in the $cb$ field could move to scales of $k < 1\,h/{\rm Mpc}$ in the halo field. Hence, if one knew the bias model one could obtain tight constraints on the neutrino mass with modern measurements of the halo field. However, the bias model parameters can have strong degeneracies with the cosmological parameters, for example, the linear bias $b_1$ is essentially degenerate with $\sigma_8$. It is thus important to marginalize over bias, and apply halo mass or number density cuts, to obtain realistic constraints.

Many of the works that compute constraints on $M_\nu$ %, 
%or find information beyond the linear power, 
are based on Fisher forecasts, for which one must take great care to avoid inaccurate results \citep{Euclid_2020, Yahia_Cherif_2021, bhandari2021fisher}. %, or by comparing a pair of phase matched simulations, which can suffer from sample variance. 
Additionally, a Fisher analysis employs asymptotic limits using the Taylor expansion of the log likelihood, which 
may not be justified in a realistic data analysis where the posteriors are often non-Gaussian. Thus, while some practitioners do go to great lengths to show that their Fisher matrices have converged,
it is unclear how credible such 
 forecasts are for higher-order statistics. 
 There is a growing trend in modern
 Statistical Inference and Machine Learning to use cross-validation as a golden standard for 
 validation of results. The same standard should 
 be adopted in cosmology as well. This means 
 setting aside some fraction of simulations 
 that are not used for training (i.e.~not used to evaluate the covariance or derivatives of summary statistics), and performing an
 end-to-end analysis on these validation 
 simulations all the way to the 
 cosmological parameters of interest, where the 
 result can be compared to the truth in terms
 of bias and variance. 
 Such an analysis is expensive, even more so if 
 the validation simulations are chosen to be produced by an 
 independent simulation code, 
 but this could be a worthwhile standard 
 validation procedure. % to avoid fake news.
 Another worthwhile verification strategy is to use null tests, in which one explicitly performs the
 analysis on setups where the signal is known to be null. An 
 example is running non-Gaussian statistical 
 analysis on Gaussian data to demonstrate that the 
 Fisher analysis does not give more 
 information than what is available in the 
 Gaussian field. 
 Thus a useful piece of future work would be to train a neural network to learn the effects of massive neutrinos on the various cosmological fields and perform all of these tests. %, perhaps by using a set of Latin hypercube simulations. 
% Nonetheless, in the absence of such work, the arguments presented in this paper provide a simple verification of constraints pertaining to neutrino mass.
In the absence of such work, we intend
for our results to give a useful rule of thumb when
proposing new statistics to measure the non-linear effects
of massive neutrinos.

\begin{acknowledgments}
We thank Francisco Villaescusa-Navaro for many insightful discussions on the manuscript. AB (Arka Banerjee) was supported by the Fermi Research Alliance, LLC under Contract No. DE-AC02-07CH11359 with
the U.S. Department of Energy, and the U.S. Department of Energy (DOE) Office of Science Distinguished Scientist Fellow Program. This material is based upon work supported by the National Science Foundation under Grant Numbers 1814370 and NSF 1839217, and by NASA under Grant Number 80NSSC18K1274.
We acknowledge the use of CAMB for computations involving the linear power spectrum \cite{CAMB}.
The \textit{Quijote} simulations \cite{quijote} can be found at \url{https://github.com/franciscovillaescusa/Quijote-simulations}. The analysis of the simulations has made use of the \textit{Pylians} libraries, publicly available at \url{https://github.com/franciscovillaescusa/Pylians3}. Some of the computing for this project was performed on the Sherlock cluster. The authors would like to thank Stanford University and the Stanford Research Computing Center for providing computational resources and support that contributed to these research results.

\end{acknowledgments}
%\newpage
\appendix

\section{Other Parameters ($\Omega_m$ and $A_s$)}
\label{app:params}

In this paper we have studied the effect of neutrino mass, $M_\nu$, on non-linear cosmic evolution. We now briefly discuss the effects of two other cosmological parameters relevant for disentangling the effects of neutrino mass from large-scale structure: $\Omega_m$ and $A_s$.

\begin{figure}[t]
\includegraphics[width=\linewidth]{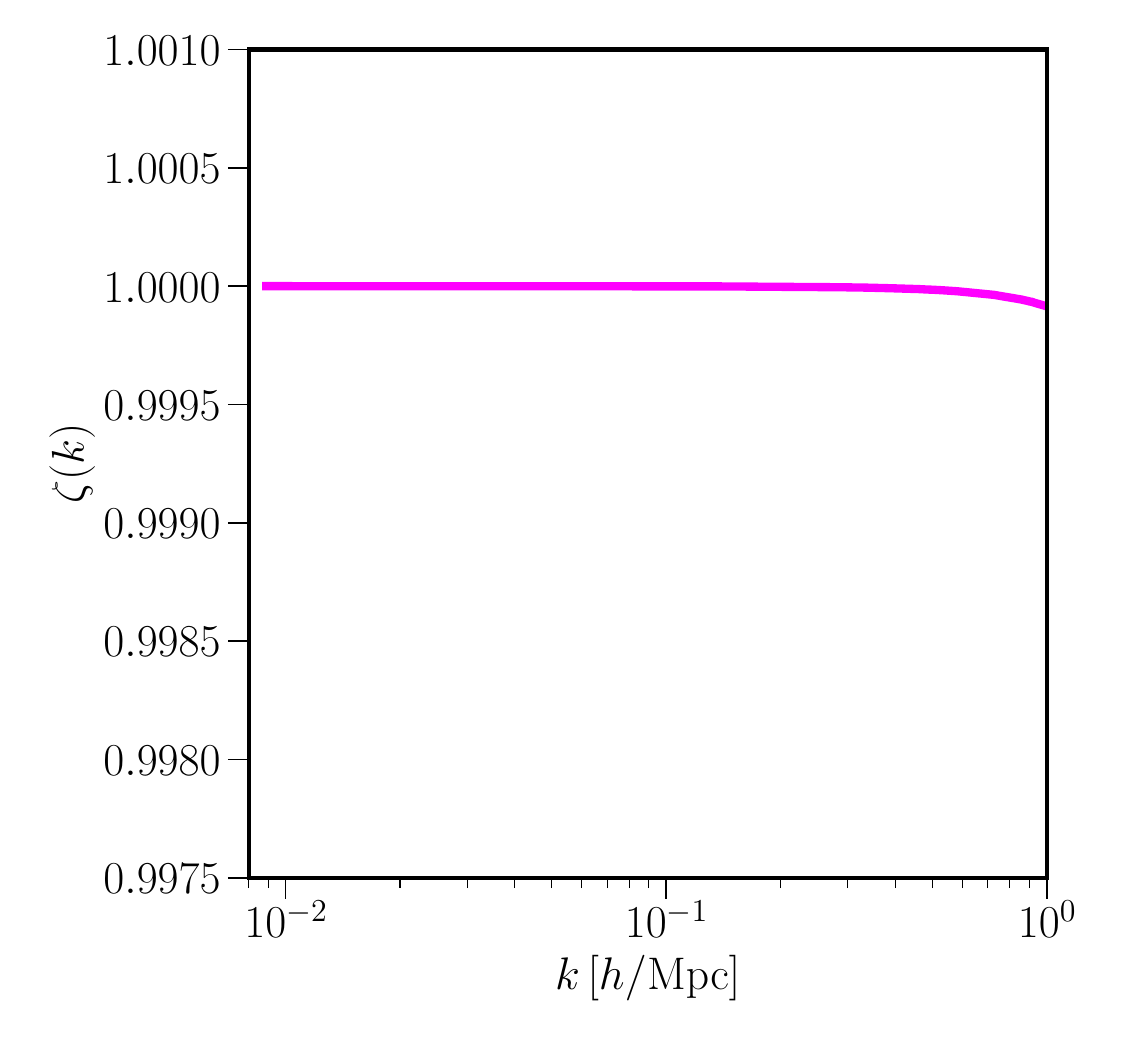}
\caption{The real part of the coherence between the $cb$ fields from two $M_\nu=0$ simulations with matched linear $P(k)$ at $z=0$, but with $\Omega_m$ differing by 10\%. It can be seen that the coherence is 1 up to $k=1\,h/{\rm Mpc}$ to $\lesssim 0.01\%$. This implies that there is negligible non-linear information regarding $\Omega_m$ in the $cb$ field. The vertical range is identical to Fig.~\ref{fig:r_cc} to enable comparison.
}
\label{fig:r_cc_Omega_m} 
\end{figure}

We first perform the real-versus-fake analysis on the cosmological parameter $\Omega_m$. We seek to 
test if non-linear evolution leaves an imprint 
at the field level. % by looking at the correlation between the two simulations.
To do so, we match the linear $P(k)$ of two $M_\nu=0$ simulations, but differ the value of $\Omega_m$ by 10\% between the two, during both backscaling and the forward $N$-body simulation. Fig.~\ref{fig:r_cc_Omega_m} shows the coherence between these two simulations, which is shown to be $\lesssim 0.01 \%$ for $k \lesssim 1\, {h/{\rm Mpc}}$. This suggests there is negligible additional information regarding $\Omega_m$ coming from the non-linear evolution that would be present 
in higher-order statistics, since the agreement 
is exact at the field level. Interestingly, this is about the same value as the coherence found for the real-versus-fake $M_\nu$ coherence found in Fig.~\ref{fig:r_cc}. 
%The little information that does arise is due to the difference in growth rate, $f$, between the two simulations, as this depends on $\Omega_m$. Also 
Note that this analysis does not take into account any change in the shape of $P(k)$ due to a change in $\Omega_m$, which is information contained in the 
initial conditions.

The other parameter of relevance when it comes to neutrino mass is the amplitude of linear fluctuations, $A_s$. As this is the amplitude of the initial linear power spectrum, it is a property of the initial conditions. Thus late-time non-linear evolution cannot produce additional information on $A_s$.

We thus conclude that there is little information regarding $M_\nu, \Omega_m$, or $A_s$ coming from non-linear cosmic evolution. Hence, for $k \lesssim 1\, {h/{\rm Mpc}}$ in the $cb$ or $\Omega m$ fields, there is no non-linear statistic that will constrain these parameters significantly better than the linear power spectrum, even after marginalizing.

\newpage   % get error when end of doc line is on new page without putting this line first
\bibliography{references, apssamp}

\end{document}